\documentclass[namedreferences]{solarphysics}
\usepackage[hyperref,optionalrh,solaromanenum]{spr-sola-addons} 
\usepackage{graphicx}     
\usepackage[shortlabels]{enumitem}
\usepackage[title]{appendix}
\usepackage{color}                       
\usepackage{breakurl}                         
\usepackage{ulem}
\usepackage{xcolor}
\usepackage{hyperref}

\begin{document}
\begin{article}
\begin{opening}

\title{A Survey of Computational Tools in Solar Physics}

\author[addressref={aff1}]{\inits{M.G.}\fnm{Monica G.}~\lnm{Bobra}\orcid{0000-0002-5662-9604}}
\author[addressref={aff2,aff3},corref,email={stuart.mumford@sheffield.ac.uk}, ]{\inits{S.}\fnm{Stuart}~\lnm{Mumford}\orcid{0000-0003-4217-4642}}
\author[addressref={aff6}]{\inits{R.}\fnm{Russell J.}~\lnm{Hewett}\orcid{0000-0001-8944-4705}}
\author[addressref={aff4}]{\inits{S.D.}\fnm{Steven D.}~\lnm{Christe}\orcid{0000-0001-6127-795X}}
\author[addressref={aff7}]{\inits{K.}\fnm{Kevin}~\lnm{Reardon}\orcid{0000-0001-8016-0001}}
\author[addressref={aff8}]{\inits{S.}\fnm{Sabrina}~\lnm{Savage}\orcid{0000-0002-6172-0517}}
\author[addressref={aff4}]{\inits{J.}\fnm{Jack}~\lnm{Ireland}\orcid{0000-0002-2019-8881}}
\author[addressref={aff9,aff10}]{\inits{T.M.D.}\fnm{Tiago M. D.}~\lnm{Pereira}\orcid{0000-0003-4747-4329}}
\author[addressref={aff11}]{\inits{B.}\fnm{Bin}~\lnm{Chen}\orcid{0000-0002-0660-3350}}
\author[addressref={aff5}]{\inits{D.}\fnm{David}~\lnm{P\'{e}rez-Su\'{a}rez}\orcid{0000-0003-0784-6909}}

\address[id=aff1]{W.W. Hansen Experimental Physics Laboratory, Stanford University, Stanford, CA 94305, USA}
\address[id=aff2]{SP$^2$RC, School of Mathematics and Statistics, The University of Sheffield, Sheffield, UK}
\address[id=aff3]{Aperio Software Ltd., Headingley Enterprise and Arts Centre, Bennett Road, Leeds, UK}
\address[id=aff6]{Department of Mathematics, Virginia Polytechnic Institute and State University, Blacksburg, VA 24061-0123, USA}
\address[id=aff4]{NASA Goddard Space Flight Center, Greenbelt, MD 20771, USA}
\address[id=aff7]{National Solar Observatory, Boulder, CO 80303, USA}
\address[id=aff8]{NASA Marshall Space Flight Center, Huntsville, AL 35812, USA}
\address[id=aff9]{Institute of Theoretical Astrophysics, University of Oslo, Oslo, Norway}
\address[id=aff10]{Rosseland Centre for Solar Physics, University of Oslo, Oslo, Norway}
\address[id=aff11]{Center for Solar-Terrestrial Research, New Jersey Institute of Technology, Newark, NJ 07102-1982, USA}
\address[id=aff5]{University College London, London, UK}

\runningauthor{Bobra M.G. {\it et al.}}
\runningtitle{A Survey of Computational Tools in Solar Physics}

\begin{abstract}
The SunPy Project developed a 13-question survey to understand the software and hardware usage of the solar physics community. 364 members of the solar physics community, across 35 countries, responded to our survey. We found that 99$\pm$0.5\% of respondents use software in their research and 66\% use the Python scientific software stack. Students are twice as likely as faculty, staff scientists, and researchers to use Python rather than Interactive Data Language (IDL). In this respect, the astrophysics and solar physics communities differ widely: 78\% of solar physics faculty, staff scientists, and researchers in our sample uses IDL, compared with 44\% of astrophysics faculty and scientists sampled by \citet{Momcheva}. 63$\pm$4\% of respondents have not taken any computer-science courses at an undergraduate or graduate level. We also found that most respondents utilize consumer hardware to run software for solar-physics research. Although 82\% of respondents work with data from space-based or ground-based missions, some of which ({\it e.g.} the {\it Solar Dynamics Observatory} and {\it Daniel K. Inouye Solar Telescope}) produce terabytes of data a day, 14\% use a regional or national cluster, 5\% use a commercial cloud provider, and 29\% use exclusively a laptop or desktop. Finally, we found that 73$\pm$4\% of respondents cite scientific software in their research, although only 42$\pm$3\% do so routinely.
\end{abstract}
\keywords{Instrumentation and Data Management}
\end{opening}

\section{Introduction}
\label{S-Introduction} 
The SunPy Project \citep{sunpyapj} facilitates and promotes the use and development of community-led, free, and open source\footnote{According to the Open Source Initiative, stewards of the Open Source Definition (available at {\textsf{\href{https://opensource.org/osd}{opensource.org/osd}}}), open source software consists of source code under an open source license. Open source licenses “allow modifications and derived works, and must allow them to be distributed under the same terms as the license of the original software.” In addition, open source software must not discriminate against persons, groups, or fields and the associated licenses must be non-specific, non-restrictive, and technology-neutral.} data-analysis software for solar physics based on the scientific Python environment. To better understand the software and hardware preferences of the solar-physics community, the Project developed a 13-question survey (reproduced in Appendix \ref{A-SurveyQs}) and disseminated it internationally\footnote{The UK Solar Physics, European Physical Society's Solar Physics Division, American Astronomical Society's Solar Physics Division, the solar-physics subdivision of the Astronomical Society of Japan, the Astronomical Society of India, and the Brazilian Astronomical Society organizations advertised the survey to their members. The SunPy Project also advertised the survey on the \textsf{@SunPyProject} Twitter account and sent it to the \textsf{sunpy} and \textsf{sunpy-dev} e-mail lists, both of which are public.} over a six-month period between 7 February 2019 and 28 July 2019. 

Many of the survey questions were similar (and in some cases, identical) to those posed by \citet{Momcheva} in an informal survey of 1142 members of the astrophysics community. The SunPy Project did this deliberately to compare software preferences between the solar and astrophysics communities.

This article presents the survey results, derived from analyzing 364 responses from community members across 35 countries.
All of the survey responses, along with the code \citep{pandas101, matplotlib313, seaborn0100, numpy, survey020} to analyze these data and produce the figures in this article, are publicly available at {\textsf{\href{https://github.com/sunpy/survey}{github.com/sunpy/survey}}}.

\section{Demographics}
\label{S-Demographics} 

Since the SunPy Project relies largely on volunteer efforts, we chose to construct and disseminate this survey ourselves (instead of going through a formal channel such as the Statistical Research Center at the American Institute of Physics). As a result, we recognize that this survey may suffer from coverage error.

Our survey garnered 368 responses. Most of the survey respondents fit into one of four career stages: 56\% ($n$=205) described themselves as a faculty member, staff scientist, or researcher, 15\% ($n$=53) as a postdoc, 23\% ($n$=84) as an undergraduate or graduate student, and 6\% ($n$=22) as a software or instrument developer. This adds up to $n$=364. Four respondents did not fit into any career stage, and we dropped their responses from our analysis.

Community members across 35 countries\footnote{See the analysis code, available at {\textsf{\href{https://github.com/sunpy/survey}{github.com/sunpy/survey}}}, for a full list.} responded to our survey. About three-quarters of the respondents came from the US, UK, Germany, India, and Japan. Together, these five countries include about 1150 solar physicists\footnote{The Solar Physics Division of the American Astronomical Society includes 521 members (private communication, S. Savage, 28 January 2020). The UK Solar Physics community estimates “over 150 scientists" on its website, {\textsf{\href{https://www.uksolphys.org}{uksolphys.org}}}, as of 27 January 2020. The European Solar Physics Division counts 222 members (private communication, T.M.D. Pereira, 28 January 2020). The Astronomical Society of India includes approximately 100 solar physicists (private communication, D. Banerjee, 30 January 2020). The communication newsletter for the solar-physics subdivision of the Astronomical Society of Japan, called Renraku-kai, counts about 150 subscribers (private communication, K. Hayashi, 27 January 2020).}; therefore, our survey sampled roughly a quarter of the solar-physics community. Our results are based on the assumption that our sample is representative of the solar-physics community overall.

We asked respondents to identify all of the areas of research relevant to their career. Most respondents identified multiple sub-disciplines of expertise. We found that 76\% ($n$=275) work with space-based observational data, 46\% ($n$=169) work with ground-based observational data, and 26\% ($n$=93) work on building instruments. A vast majority of respondents, 82\%, work with ground-based or space-based data. 29\% ($n$=105) identified theory as a relevant sub-discipline, and 47\% ($n$=171) identified numerical simulations. 
 
Most of the survey respondents (82\%) chose to answer an optional question about whether they self-identified as an underrepresented minority; 16\% of this subset (13\% of the total sample) said yes. 79\% of respondents chose to answer another optional question about whether they self-identified as a underrepresented gender identity; 11\% of this subset (9\% of the total sample) said yes.

\section{Software Tools}
\label{S-Software}

In our survey of the solar-physics community, we found that 99$\pm$0.5\% of respondents use software in their research\footnote{While three respondents did apparently indicate that they do not use software in their research, their further answers on the survey about software package usage suggest that those might have been erroneous responses.}. In a survey of the astrophysics community, \citet{Momcheva} found that 100\% of respondents use software in their research.

\begin{figure} 
\centerline{\includegraphics[width=\textwidth,clip=]{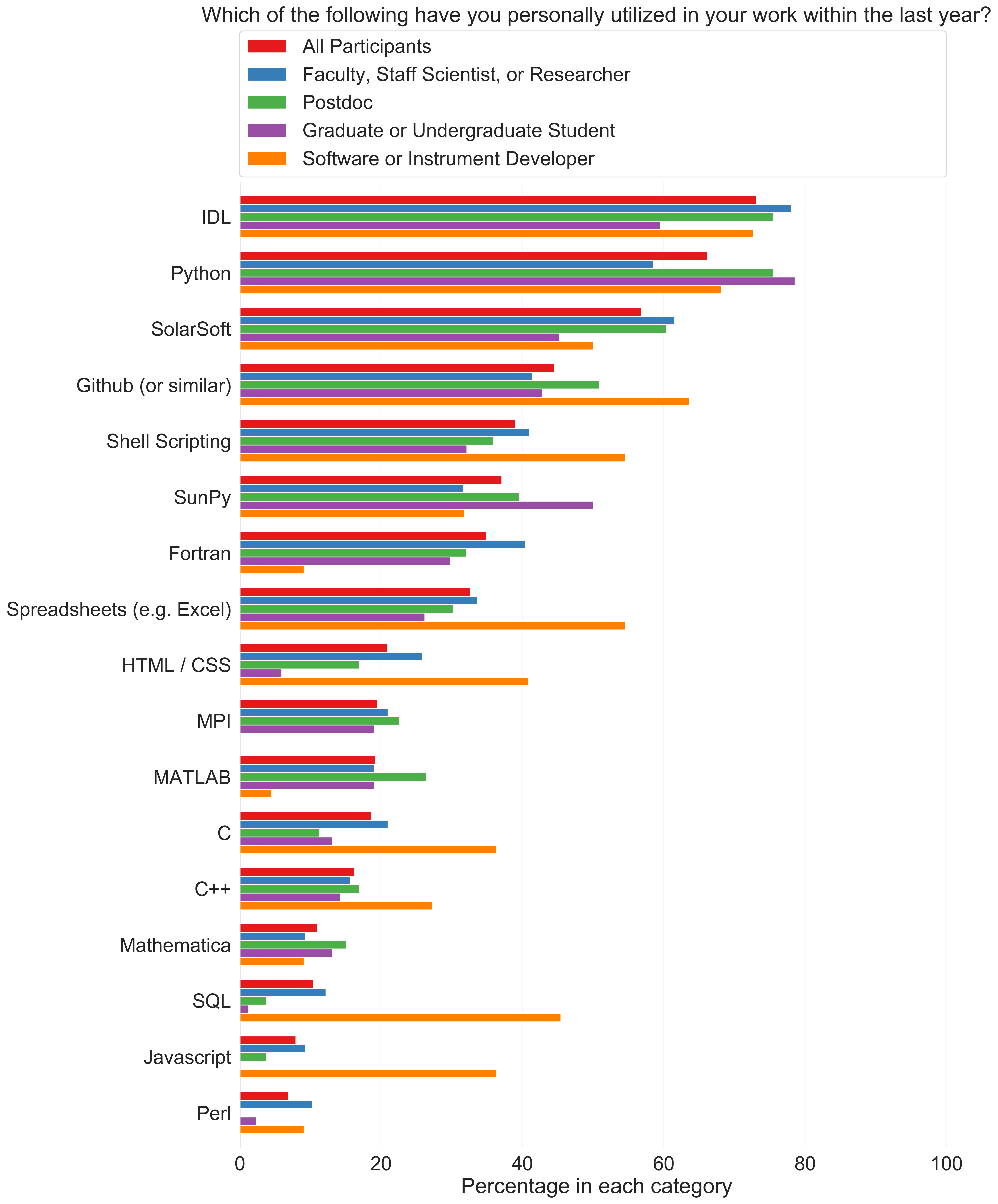}}
\caption{Summary of results for survey Question 9 ``Which of the following [software tools] have you personally utilized in your work within the
last year?''  Results are grouped by self-identified career stage (Question 2).  Respondents listed 42 different software tools; only tools used by 5\% or more of respondents are shown.}
\label{F-4a}
\end{figure}

We asked users to list all of the scientific software tools, including programming languages, software development tools, and data-analysis frameworks, that they utilized within the last year. We summarized their responses in Figure~\ref{F-4a}. We found that 66\% of respondents use the Python scientific software stack and 73\% use IDL\footnote{Where relevant, we supplied our counting error for non-demographic software and hardware related questions (Questions 6\,--\,12). For Question 6, we report $\sqrt3/364$, or 0.5\%, as the percentage error in the number of no responses. Since this question required respondents to pick one response from a binary choice, we apply that same uncertainty to the yes responses. For Questions 7, 8, 10, and 11, which required respondents to pick only one response from a list of options, we quantified the percent error in each response simply by applying the square-root rule for counting experiments \citep{taylor}. For Questions 9 and 12, which allowed respondents to select as many options as they liked, we do not calculate a percent error.}. Overall, respondents listed 42 different software tools and the average respondent used five tools in the past year.

We observe a stark contrast in usage between the two primary data-analysis languages in solar-physics research, Python and IDL, when viewed  by respondent career stage.  The earlier the career stage, the greater the percentage of Python users: 59\% of faculty, staff scientists, and researchers, 75\% of postdocs, and 79\% of students use Python. The earlier the career stage, the fewer IDL users: 78\% of faculty, staff scientists, and researchers, 75\% of postdocs, and 60\% of students use IDL. 

\begin{figure} 
\centerline{\includegraphics[width=\textwidth,clip=]{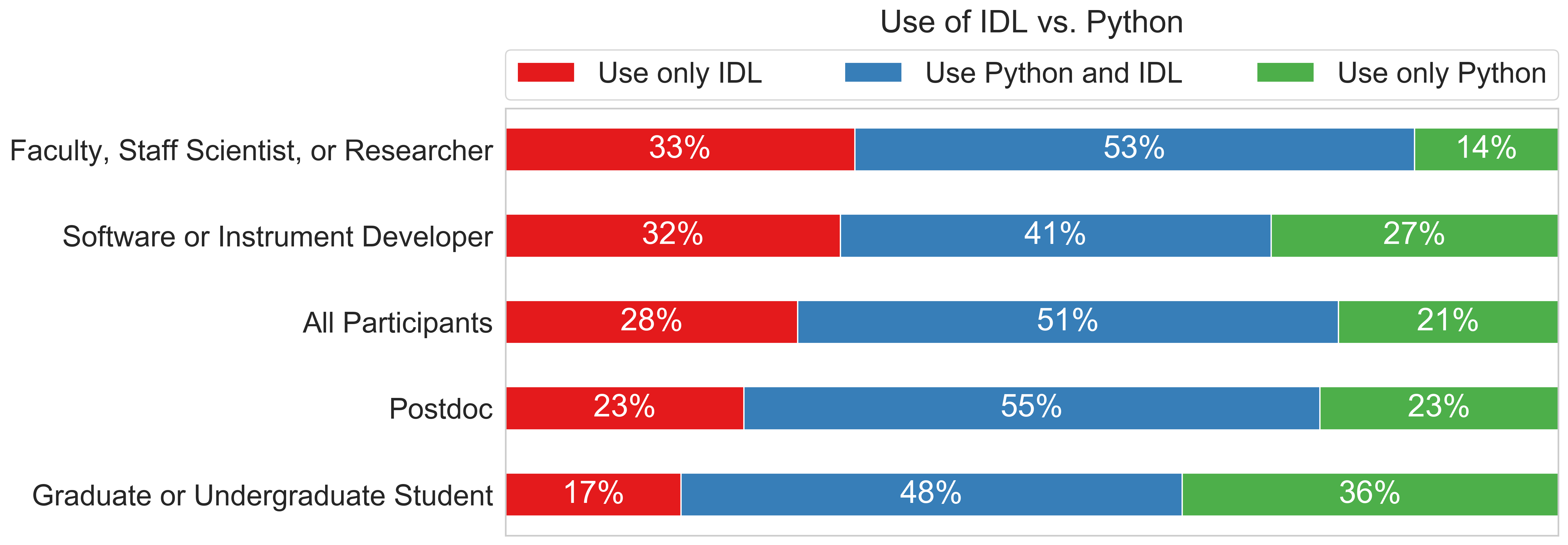}}
\caption{Comparison of respondents that report using Python or IDL exclusively by reported career role.}
\label{F-4c}
\end{figure}

Of course, these tools are not necessarily used in isolation -- about half (45\%) of respondents use both Python and IDL. Figure \ref{F-4c} shows that 28\% of respondents use IDL exclusively (in other words, they use IDL and do not use Python), while 21\% use Python exclusively. The ratio of exclusive IDL users to exclusive Python users is roughly 2:1 for faculty, staff, and research scientists and the opposite, 1:2, for students.

Figure 10 of \citet{Momcheva} shows that Python is not only the most popular programming language within their sample of the astrophysics community, but it is also the most popular within every individual career category. Our survey results show that Python is the most popular programming language only among students; IDL and Python are at parity for postdocs, and IDL is more popular than Python for faculty, staff scientists, researchers, software developers, and instrument developers. In this respect, the astrophysics and solar-physics communities differ widely: 78\% of solar-physics faculty, staff scientists, and researchers in our sample use IDL\footnote{The use of IDL by the solar-physics community may be explained partly by how instrument teams provide their data.  Many instrument teams provide data that have been calibrated to a low level, plus software that allows the data to be further calibrated for scientific use.  The advantage of this model of scientific-data provision is that as knowledge of the instrument improves over time, the software can be updated to provide better high-level science-ready data products.  A side-effect of this model of scientific-data provision is that scientific use of the data requires use of a particular package/language. Since many instrument teams chose to take advantage of the significant functionality provided by the SolarSoftWare (SSW: \citealt{Freeland:1998we}) package, much of the software required to create higher-level data products is written in the primary language of SSW: IDL.  Hence the model of scientific-data provision may explain why IDL is used by a significant proportion of respondents.}, compared with 44\% of astrophysics faculty and scientists sampled by \citet{Momcheva}.

\begin{figure}
\centerline{\includegraphics[width=\textwidth,clip=]{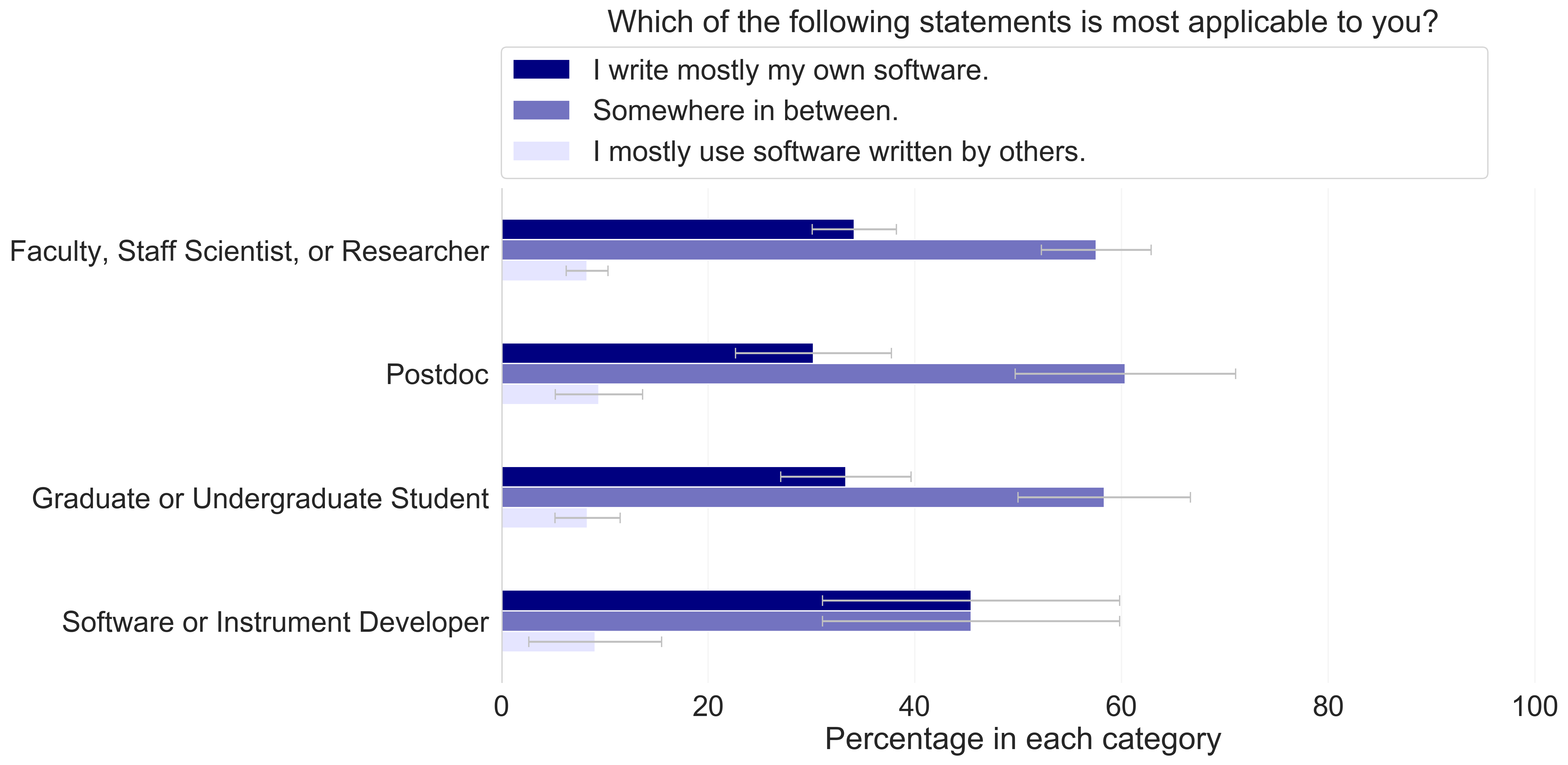}}
\caption{Comparison of respondent's software development and use activities by reported career role, with uncertainty estimate.}
\label{F-3}
\end{figure}

The two groups of respondents share the same statistics, however, when it comes to writing software. In both the astrophysics and solar-physics communities, roughly a third of respondents write their own software most of the time (see Figure \ref{F-3} of this article and Figure 3 of \citealp{Momcheva}). Furthermore, about 90\% of respondents in both communities often or occasionally write their own software (see the same figures).

\section{Education and Training}
\label{S-Education} 
\begin{figure}
\centerline{\includegraphics[width=\textwidth,clip=]{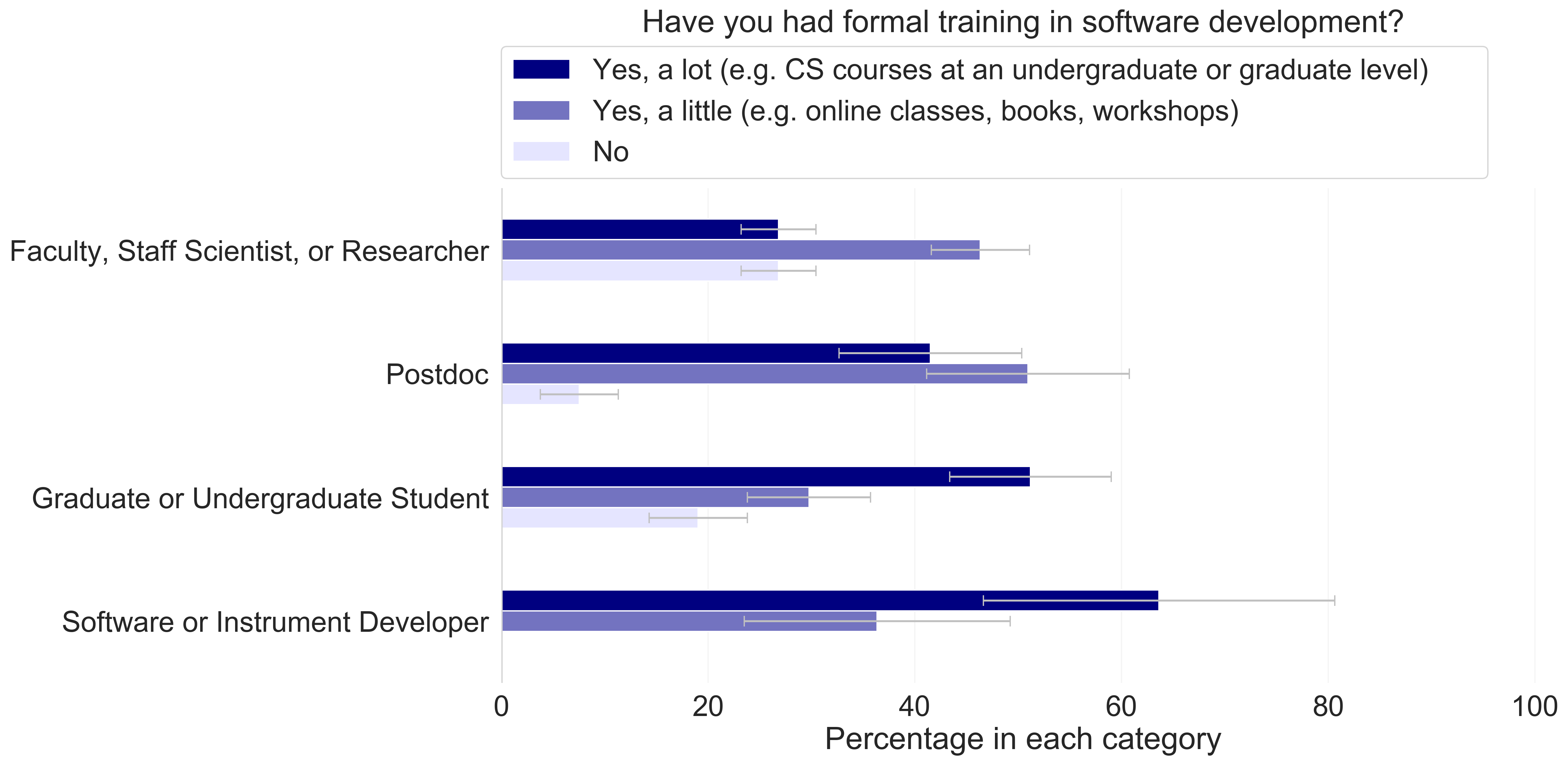}}
\caption{Comparison of respondent's formal computer-science education activities (at both undergraduate and graduate level) by reported career role, with uncertainty estimate.}
\label{F-2}
\end{figure}

Although 99$\pm$0.5\% of respondents use software in their research and 91$\pm$5\% often or occasionally write their own software, 63$\pm$4\% of respondents have not had any formal training ({\it e.g.} computer-science courses) at an undergraduate or graduate level. We found that people who write mostly their own software are no better trained than everyone else: 44$\pm$6\% of people who write their own software reported ``a lot ({\it e.g.} computer science courses)" of formal training, compared with 37$\pm$3\% overall. We also found that students today are twice as likely to have a lot of formal training in programming compared with faculty, researchers, and staff scientists (see Figure \ref{F-2}). The amount of training does not vary with area of expertise; each sub-discipline shows roughly the same amount of formal training as the general population (37$\pm$3\%).

\section{Hardware Tools}
\label{S-Hardware} 
\begin{figure}
\centerline{\includegraphics[width=\textwidth,clip=]{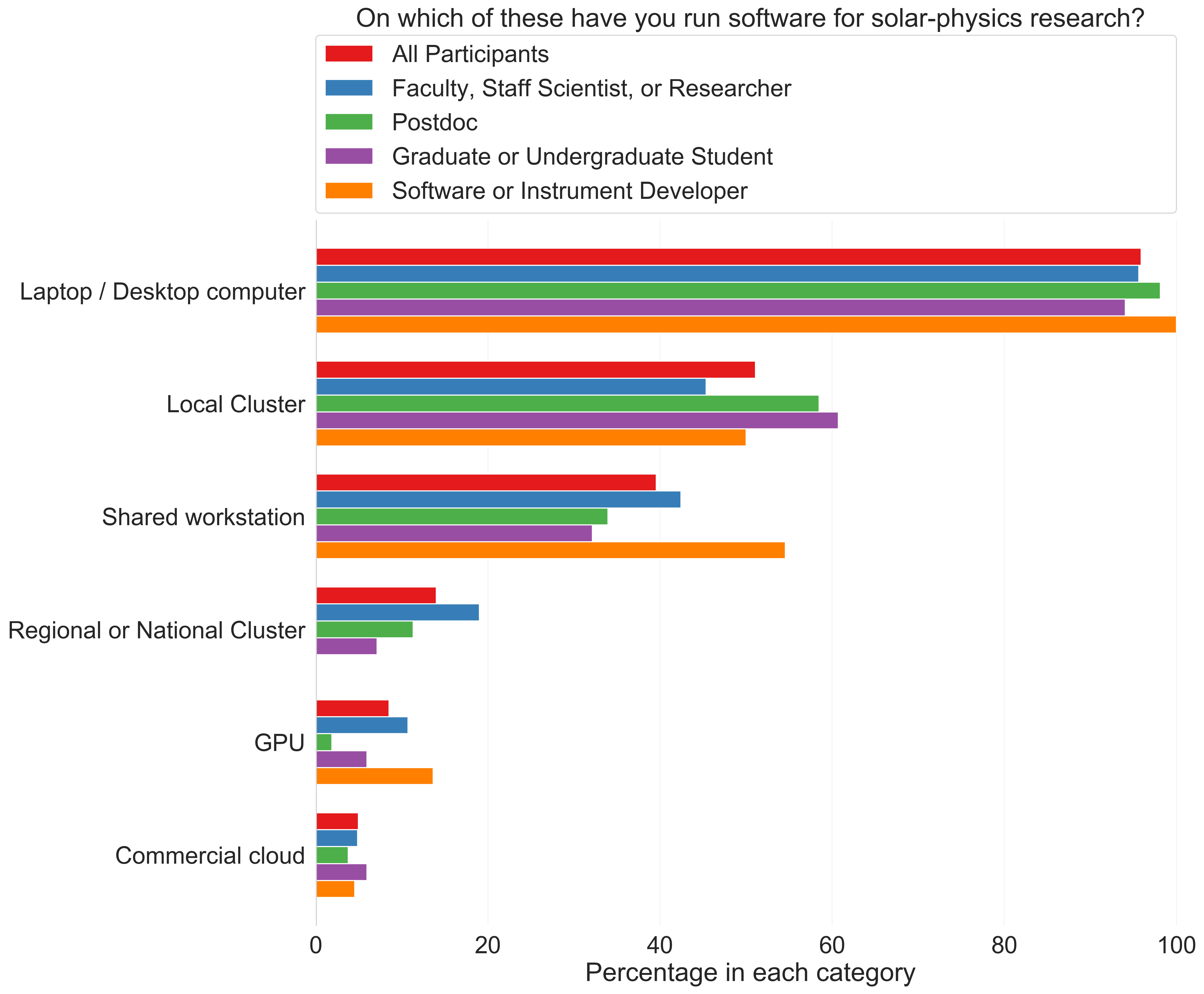}}
\caption{Responses to Question 12, related to computer resource and hardware usage, broken down by career role (Question 2).}
\label{F-7a}
\end{figure}

We also found that most respondents utilize consumer hardware to run software for solar-physics research. Although 82\% of respondents work with space-based or ground-based data, and some of these missions ({\it e.g.} the {\it Solar Dynamics Observatory} and {\it Daniel K. Inouye Solar Telescope}) produce terabytes of data per day, 14\% use a regional or national cluster\footnote{We recognize that some countries, such as the United States, require citizenship or permanent residence status to use these clusters.} and 5\% use a commercial cloud provider (see Figure \ref{F-7a}). 29\% use exclusively a laptop or desktop. The community puts considerable effort into maintaining clusters and workstations, with 40\% of respondents using a shared workstation, 51\% using a local cluster, and 96\% using a laptop or desktop.

These percentages vary significantly by sub-discipline. A larger percentage of respondents in the numerical simulations and theory sub-disciplines use local clusters (63\% and 60\%, respectively, compared with 51\% overall) and regional or national clusters (26\% and 26\%, respectively, compared with 14\% overall).

\begin{figure}
\centerline{\includegraphics[width=\textwidth,clip=]{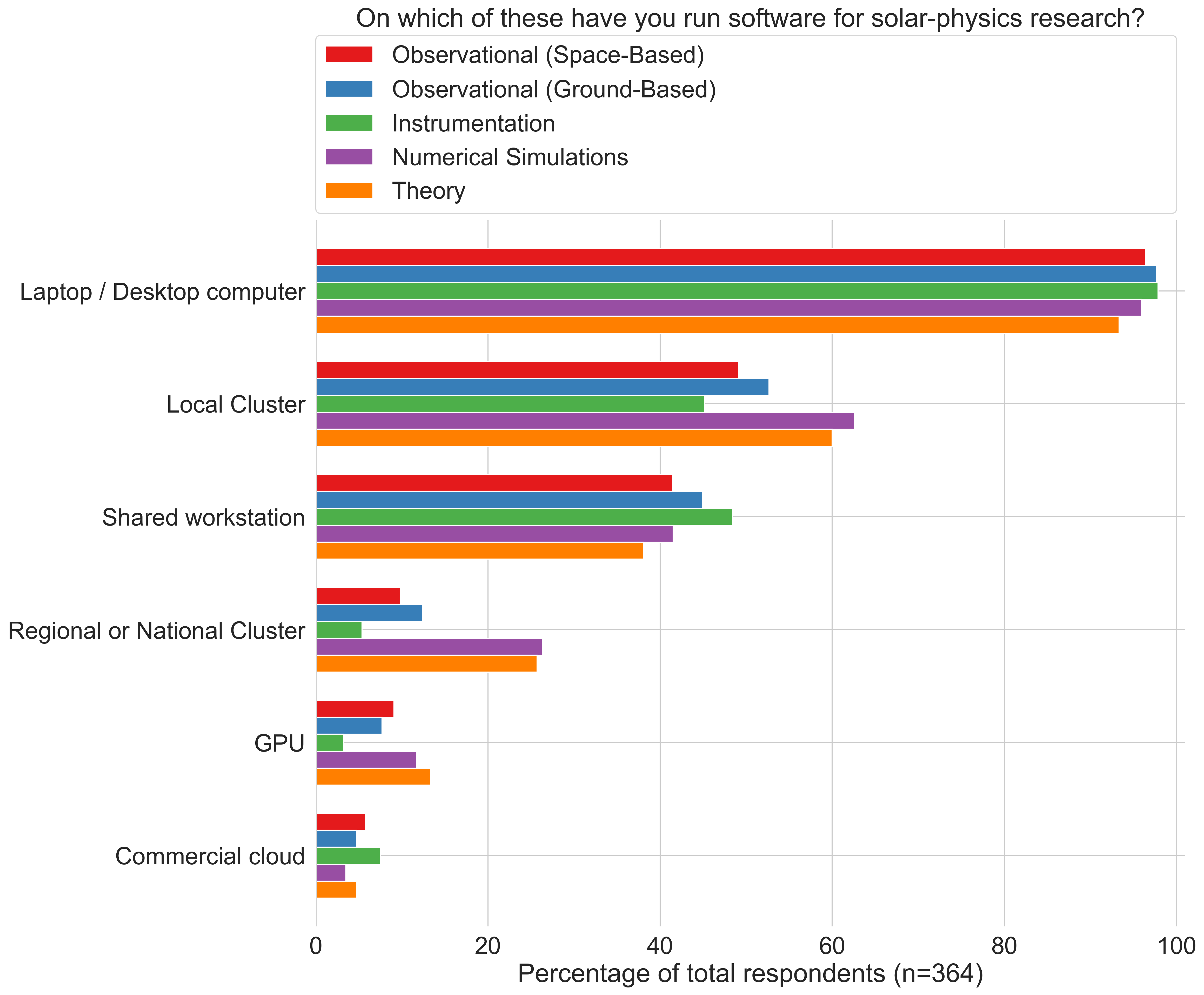}}
\caption{Responses to Question 12, related to computer resource and hardware usage, broken down by solar-physics research area (Question 1).}
\label{F-7b}
\end{figure} 

\section{Citing Scientific Software}
\label{S-Citations} 
\begin{figure}
\centerline{\includegraphics[width=0.7\textwidth,clip=]{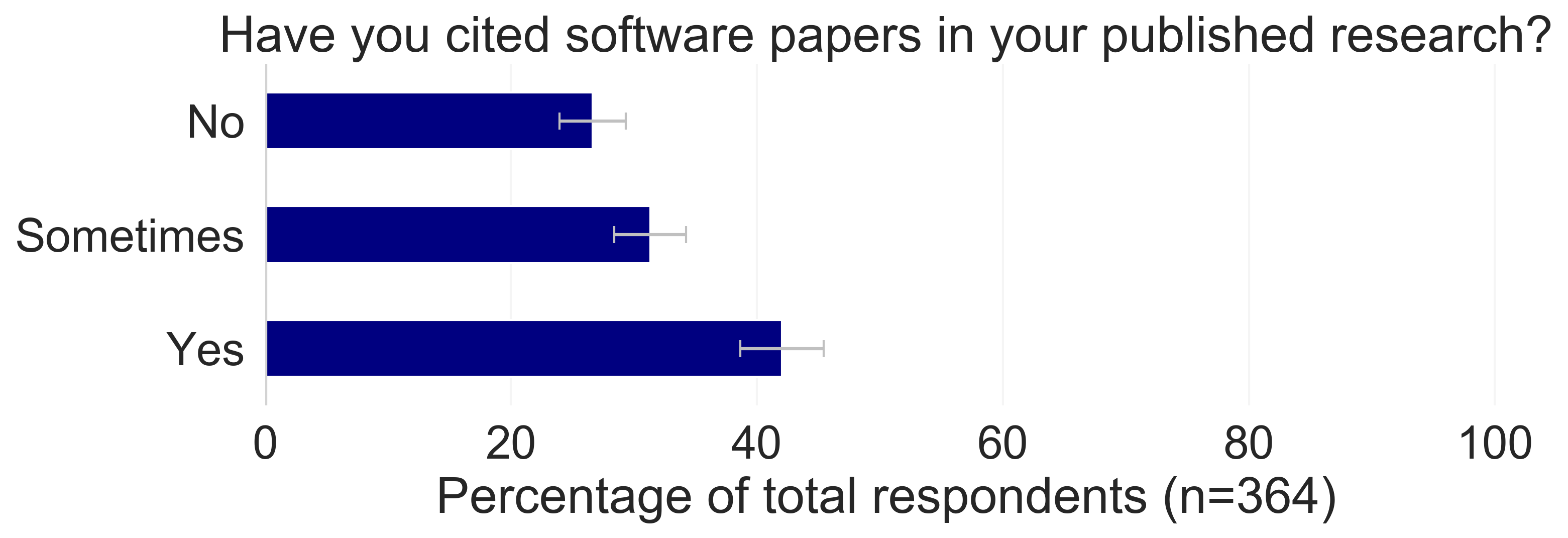}}
\caption{Responses to Question 10, ``Have you cited software papers in your published research?''}
\label{F-5}
\end{figure}

\begin{figure}
\centerline{\includegraphics[width=\textwidth,clip=]{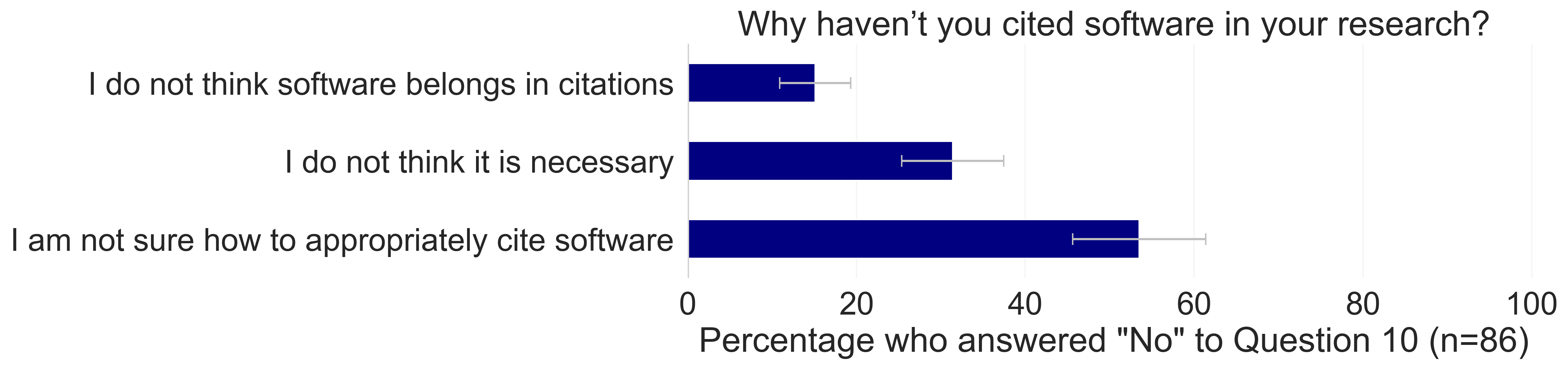}}
\caption{Responses to Question 11, ``Why haven't you cited software in your research?'', for those that responded ``No'' to Question 10.}
\label{F-6}
\end{figure}

Figure \ref{F-5} shows that 73$\pm$4\% of respondents cite scientific software in their research, although only 42$\pm$3\% do so routinely. Roughly a quarter (27$\pm$3\%) never cite scientific software in their research. When asked why, about half (53$\pm$8\%) responded that they do not know how to appropriately cite scientific software (see Figure \ref{F-6}); we note that only 4$\pm$1\% of respondents do not think software belongs in citations.

\section{Discussion}
\label{S-Discussion} 

Scientific software is an indispensable component of the modern scientific research workflow \citep{rude}. Virtually all of the solar-physics community uses software in their research. Based on this fact, we find three of the statistics presented in this article worrisome. First, similar to the astrophysics community\footnote{\citet{Momcheva} found that only 8$\pm$1\% of the astrophysics community received substantial training; however, their question did not define “a lot" or “a little".}, a significant fraction of the solar-physics community (63$\pm$4\% of respondents) have not taken any computer-science courses at an undergraduate or graduate level. Second, most of the solar-physics community (82\% of respondents) works with space-based or ground-based facilities, several of which produce terabyte- or petabyte-sized data sets, and nearly a third of the community (29\% of respondents) uses exclusively a laptop or desktop to run software for solar-physics research. It is unclear whether the computing power offered by laptops and desktops limit the type of scientific endeavors in solar-physics. Finally, less than half of the community (42$\pm$3\% of respondents) routinely cites scientific software in their research.

The United States \cite{NAP2018} report entitled {\it Software Policy Options for NASA Earth and Space Sciences} recognizes the lack of education in software development among scientists. The report recommends initiating and sponsoring “programs to educate and train researchers in open source best practices," suggesting topics such as “export controls, licensing and intellectual property, workflows, and software development.” This includes sponsoring community members to attend conferences about open source software development, such as Python in Astronomy ({\textsf{\href{https://openastronomy.org/pyastro}{openastronomy.org/pyastro}}}) or Scientific Computing with Python ({\textsf{\href{https://conference.scipy.org}{conference.scipy.org}}}), take online courses about software development, available on learning platforms such as Coursera ({\textsf{\href{https://www.coursera.org}{coursera.org}}}) and edX ({\textsf{\href{https://www.edx.org}{edx.org}}}), join workshops like those led by The Carpentries ({\textsf{\href{https://carpentries.org}{carpentries.org}}}), and develop training programs, such as the Large Synoptic Survey Telescope's Data Science Fellowship program ({\textsf{\href{https://astrodatascience.org}{astrodatascience.org}}}). Our findings in Section \ref{S-Education} show that the solar physics community could benefit immensely from education and training in open source software.

The Ford Foundation's report, entitled {\it Roads and Bridges: The Unseen Labor Behind Our Digital Infrastructure} \citep{fordfoundation}, also suggests “expanding the pool of contributors so that more people, and more types of people, can build and sustain public software together." Increasing the diversity of the talent pool, which is still lacking in the solar-physics community, will help sustain a long-term future for open source software in solar-physics.

However, maximizing the scientific return of large data sets, such as those produced by the Solar Dynamics Observatory and the Daniel K. Inouye Solar Telescope, requires both skill in software development and computational resources. The United States \citet{NAP2020} report entitled {\it Progress Toward Implementation of the 2013 Decadal Survey for Solar and Space Physics: A Midterm Assessment} and the Kavli Foundation series of workshops called {\it Petabytes To Science} \citep{petabytes} recommend adopting science platforms, which co-locate both data and computational resources required to analyze these data. In this paradigm, users run software in an external computing environment where the data lives, instead of moving the data to a desktop or laptop where the software lives. The astrophysics community already developed several science platforms, such as the ASTRO Data Lab ({\textsf{\href{https://datalab.noao.edu}{datalab.noao.edu}}}), run by the NSF’s National Optical-Infrared Astronomy Research Laboratory. We encourage the solar-physics community to fund the development of science platforms so that scientists are not restricted by the computational power of consumer hardware for analyses involving terabytes of data.

Finally, we recognize that software development, and hardware development, takes a vast amount of time. This time is rarely recognized by the academic community, which largely rewards publications. Therefore, we encourage the community to publish scientific software (by submitting articles that describe research software to refereed journals and archiving this software in publicly available digital repositories; see {\textsf{\href{https://guides.github.com/activities/citable-code}{guides.github.com/activities/citable-code}}}), cite scientific software (see Appendix \ref{A-Cite} about how to cite scientific software), and \textit{count} scientific software as a co-equal research artifact when considering career evaluation. This has two benefits: it gives academic credit and career recognition to those who write software and it makes it easier to reproduce studies in solar-physics.

Some of the earliest advocates for scientific reproducibility, \citet{claerbout} and \citet{Buckheit95wavelaband}, suggested that a journal article “about computational science in a scientific publication is not the scholarship itself, it is merely advertising of the scholarship.” The actual scholarship, they argue, is the code and development environment used to generate the results. Preserving these elements of scholarship require tools like version control, which create snapshots of software or data as they change over time. At the moment, less than half the community (44\% of respondents) uses version control\footnote{We found that 44\% of respondents selected the option “Github (or similar)" in Question 9. However, we realize this option is ambiguous. In retrospect, we should have provided “Git, Github, or similar" instead of “Github (or similar)" as an option in Question 9.}. The United States \cite{NAP2019} report entitled {\it Reproducibility and Replicability in Science} recommends that “researchers should convey clear, specific, and complete information about any computational methods and data products that support their published results in order to enable other researchers to repeat the analysis," including the data, study methods, and computational environment.

Scientists make a critical choice when selecting a computational environment, because the quality of our tools informs the quality of our research. A large fraction of the community uses the Python scientific-software stack (66\% of respondents). This number will only grow over time, since Python is the most popular programming language among students in the solar-physics community (79\% of students who took our survey use Python).

There are a number of reasons why the Python scientific-software stack is growing in prominence both in the solar-physics community and many other scientific disciplines\footnote{The number of contributors to the SunPy codebase grew by an average rate of one per month since 2011 \citep[see][Figure 1]{sunpyapj}. According to the 2019 Stack Overflow developer survey, Python is the fastest-growing major programming language today (see {\textsf{\href{https://insights.stackoverflow.com/survey/2019}{insights.stackoverflow.com/survey/2019}}}); furthermore, most universities use Python to teach computer science \citep{guo2014}.}. Interoperability between many packages for numerical methods, plotting, astronomy, statistics, and computing ({\it e.g.} \citealp{scipy, numpy, pandas, matplotlib, astropy2018, astroML, scikit-learn, dask}) allows researchers to write code with relative speed and ease. The rise of more than fifty packages in heliophysics alone (see {\textsf{\href{http://heliopython.org}{heliopython.org}}})  enables interdisciplinary analysis across traditionally isolated fields. The open-development model\footnote{An open-development model goes beyond providing open source software, it also includes making project-level decisions in publicly-visible and accessible spaces, such as mailing lists, and inviting input from the user and developer communities \citep{2019BAAS...51g.180T}.}, adopted by most of the scientific Python ecosystem, improves the longevity of software since anyone can contribute to the codebase and no single institution or person controls the software.

For these reasons, the United States \cite{NAP2018} report entitled {\it Software Policy Options for NASA Earth and Space Sciences} recommends that the “NASA Science Mission Directorate should explicitly recognize the scientific value of open source software and incentivize its development and support, with the goal that open source science software becomes routine scientific practice.” As the SunPy Advisory Board, we endorse this recommendation not only for the NASA Science Mission Directorate but for scientific funding agencies worldwide. 

\begin{acks}
We would like to thank Ivelina Momcheva, Erik Tollerud, and Nabil Freij for their help and guidance on this project, the numerous professional astronomy societies who helped publicize this survey, and everyone who took this survey.
\end{acks}

\section*{Disclosure of Potential Conflicts of Interest} The authors, all members of the SunPy Advisory Board serving in a volunteer, unpaid, capacity, declare that they have no conflicts of interest.

\begin{appendices}
\section{Survey Questions}
\label{A-SurveyQs}
The full contents of the survey, distributed as a Google Form, appear below. All the responses to the Question 13, an optional question which solicited general, free-form comments, are publicly available at {\textsf{\href{https://github.com/sunpy/survey}{github.com/sunpy/survey}}}.

\begin{enumerate}[1.]
\item Which of these areas of solar physics do you work in? Check all that apply.
    \begin{itemize}
    \item[$\square$] Observational (Space-Based)
    \item[$\square$] Observational (Ground-Based)
    \item[$\square$] Numerical Simulations
    \item[$\square$] Theory
    \item[$\square$] Instrumentation
    \end{itemize}
 
\item How would you describe the stage of your career?
    \begin{itemize}
    \item[$\ocircle$] Undergraduate student
    \item[$\ocircle$] Graduate student
    \item[$\ocircle$] Postdoc
    \item[$\ocircle$] Faculty, Staff Scientist, Researcher
    \item[$\ocircle$] Software Developer
    \item[$\ocircle$] Instrument Developer
    \item[$\ocircle$] Retired
    \item[$\ocircle$] My role is something other than solar physics or software development
    \item[$\ocircle$] Other (Respondents can enter their own description.)
    \end{itemize}

\item What country is your institution in? (Respondents check appropriate country from a list of options.)

\item Do you self-identify as one or more underrepresented minorities in solar physics? This question is optional.
    \begin{itemize}
    \item[$\ocircle$] Yes
    \item[$\ocircle$] No
    \end{itemize}

\item Do you self-identify as a unrepresented gender identity in Solar Physics? This question is optional.
    \begin{itemize}
    \item[$\ocircle$] Yes
    \item[$\ocircle$] No
    \end{itemize}

\item Do you use software in your research?
    \begin{itemize}
    \item[$\ocircle$] Yes
    \item[$\ocircle$] No
    \end{itemize}

\item Have you had formal training in programming?
    \begin{itemize}
    \item[$\ocircle$] Yes, a lot (e.g. CS courses at an undergraduate or graduate level)
    \item[$\ocircle$] Yes, a little (e.g. online classes, books, workshops)
    \item[$\ocircle$] No
    \end{itemize}

\item Which of the following statements is most applicable to you?
    \begin{itemize}
    \item[$\ocircle$] I write mostly my own software.
    \item[$\ocircle$] I mostly use software written by others.
    \item[$\ocircle$] Somewhere in between.
    \end{itemize}

\item Which of the following have you personally utilized in your work within the last year? Check all that apply.
    \begin{itemize}
    \item[$\square$] IDL
    \item[$\square$] SolarSoft
    \item[$\square$] Python
    \item[$\square$] SunPy
    \item[$\square$] Shell Scripting
    \item[$\square$] C
    \item[$\square$] C++
    \item[$\square$] Fortran
    \item[$\square$] IRAF
    \item[$\square$] Perl
    \item[$\square$] Javascript
    \item[$\square$] Julia
    \item[$\square$] MATLAB
    \item[$\square$] Java
    \item[$\square$] R
    \item[$\square$] SQL
    \item[$\square$] Ruby
    \item[$\square$] HTML/CSS
    \item[$\square$] Spreadsheets (e.g. Excel)
    \item[$\square$] Mathematica
    \item[$\square$] MPI
    \item[$\square$] Github (or similar)
    \item[$\square$] Other (Respondents can enter their own description.)
    \end{itemize}

\item Have you cited software papers in your published research?
    \begin{itemize}
    \item[$\ocircle$] Yes
    \item[$\ocircle$] Sometimes
    \item[$\ocircle$] No
    \end{itemize}

\item If ‘No’ for the previous question: Why haven’t you cited software in your research?
    \begin{itemize}
    \item[$\ocircle$] I am not sure how to appropriately cite software
    \item[$\ocircle$] I do not think it is necessary
    \item[$\ocircle$] I do not think software belongs in citations
    \end{itemize}

\item On which of these have you run software for solar-physics research?
    \begin{itemize}
    \item[$\square$] Laptop / Desktop computer
    \item[$\square$] Shared workstation
    \item[$\square$] Local Cluster
    \item[$\square$] Regional or National Cluster
    \item[$\square$] GPU
    \item[$\square$] Commercial cloud
    \end{itemize}

\item Do you have any comments? (This is a free form response; comments are not required. Please feel free to give us feedback about topics like: version control, collaborative coding platforms such as Github, standard or best practices in coding, operating systems, text editors, or your personal experience with writing code and releasing software, or general thoughts about SunPy).
\end{enumerate}

\section{Citing Scientific Software}
\label{A-Cite}
To cite scientific software, please follow these two steps:

\begin{enumerate}[1.]
\item {\bf Cite the refereed journal article describing the research software.} To find this article, visit the website for a software package and look for citation instructions. For example, the SunPy website includes dedicated citation instructions and an associated BibTex entry.

\item {\bf Cite the software archive.} Publicly available digital repositories, such as Zenodo, issue a Digital Object Identifier (DOI) for archived software. (Some institutions also provide digital repositories as part of their library system.) Generally, open source software projects in the Python scientific stack will archive their software every time that they release a new version. For example, the SunPy Github page includes the Zenodo DOI for the most recent release (as of this writing, v1.1.1); clicking on it leads to the Zenodo deposit, which provides an associated BibTex entry.
\end{enumerate}

Many projects release multiple versions of software per year, but they only write a refereed journal article once in a while (for example, the SunPy Project published an article about the v1.0 release, but they will not publish an article about the v1.1.1 release). Therefore, creating reproducible results requires citing both the journal article and the software archive. Here is an example: This research used version 1.1.1 \citep{sunpyv111} of the SunPy open source software package \citep{sunpyapj}.

\end{appendices}

\bibliographystyle{spr-mp-sola}
\bibliography{main}  

\end{article} 
\end{document}